\documentclass[10pt, conference]{IEEEtran}
\IEEEoverridecommandlockouts
\usepackage{enumitem}
\usepackage[nocompress]{cite}

\usepackage[utf8]{inputenc}
\usepackage{textcmds}
\usepackage{hyperref}

\usepackage{multirow}
\usepackage[table]{xcolor}

\usepackage{tabularx}
\usepackage[hooks, breakable]{tcolorbox}
\usepackage{wrapfig}
\def\BibTeX{{\rm B\kern-.05em{\sc i\kern-.025em b}\kern-.08em
    T\kern-.1667em\lower.7ex\hbox{E}\kern-.125emX}}

\definecolor{rowcol}{HTML}{f5f2e7}

\definecolor{quotecolor}{HTML}{5b2c6f}

\definecolor{patterncolor}{HTML}{D5A021}
\definecolor{causecolor}{HTML}{44AF69}
\definecolor{recommendcolor}{HTML}{048ba8}

\newcommand{\patternbox}[1]{
\textcolor{patterncolor}{#1}}

\newcommand{\causebox}[1]{
\textcolor{causecolor}{#1}}

\newcommand{\quotebox}[2]{
\begin{tcolorbox}[size=small, rightrule=3mm,
colback=recommendcolor!10!white, colframe=quotecolor!30!white, breakable, ]
\small``#1"$_{#2}$
\end{tcolorbox} \normalsize}

\newcommand{\simplequote}[2]{
\textit{\textcolor{quotecolor}{``#1"}$_{#2}$}
}

\newcommand{\symptom}[1]{
\indent \textcolor{patterncolor}{#1}:
}

\newcommand{\cause}[1]{
\indent \textcolor{causecolor}{#1}:
}

\newcommand{\recommendation}[1]{
\indent \textcolor{recommendcolor}{#1}:
}

\newcommand{\pattern}[1]{
\subsection{#1}
}

\author{\IEEEauthorblockN{Alina Mailach}
    \IEEEauthorblockA{
    \textit{Leipzig University}\\
    \textit{ScaDS.AI Dresden/Leipzig}
    }
\and
\IEEEauthorblockN{Norbert Siegmund}
    \IEEEauthorblockA{
    \textit{Leipzig University}\\
    \textit{ScaDS.AI Dresden/Leipzig}
    }
    
}

\usepackage{comment}

\begin{document}

\title{Socio-Technical Anti-Patterns in Building ML-Enabled Software\\ {\large Insights from Leaders on the Forefront}}
\maketitle

\begin{abstract}
Although machine learning (ML)-enabled software systems seem to be a success story considering their rise in economic power, there are consistent reports from companies and practitioners struggling to bring ML models into production. Many papers have focused on specific, and purely technical aspects, such as testing and pipelines, but only few on socio-technical aspects.

Driven by numerous anecdotes and reports from practitioners, our goal is to collect and analyze socio-technical challenges of productionizing ML models centered around and within teams.
To this end, we conducted the largest qualitative empirical study in this area, involving the manual analysis of 66 hours of talks that have been recorded by the MLOps community. 

By analyzing talks from practitioners for practitioners of a community with over 11,000 members in their Slack workspace, we found 17 anti-patterns, often rooted in organizational or management problems. We further list recommendations to overcome these problems, ranging from technical solutions over guidelines to organizational restructuring.
Finally, we contextualize our findings with previous research, confirming existing results, validating our own, and highlighting new insights.
\end{abstract}

\section{Introduction}

\quotebox{So many guests have come on [at MLOps Community Channel] and said that MLOps is an organizational problem. It's not a technology problem.} {M46}

In just a few years, machine learning (ML)-enabled software systems became ubiquitous in our daily lives. The huge number of existing applications empowered with cognitive and visual AI capabilities and the emergence of entirely new business domains based on ML draw an indisputable success story. Yet, we see only the fraction of products that made it from development to production~\cite{venturebeat, gartner}.

There is a large body of work providing explanations to this productionization challenge of ML models, such as technical debt during development~\cite{Sculley2015}, automation and pipeline challenges in the area of MLOps~\cite{GrKo+21,Muralidhar2021},  and testing and debugging~\cite{SWR+18, TPJ+18, MLL+18}. However, often these works are focusing on a pure technical level. This is surprising as building ML systems is usually a multi-person, or even a multi-team project~\cite{Kreuzberger2022}. Gail Murphy framed the lack of such studies in her recent keynote ``Is software engineering research addressing software engineering problems?" at ASE'20~\cite{murphy}.

Only few papers address the socio-technical challenges. For example, Granlund and others have found profound organizational challenges when integrating and scaling ML applications across organizations~\cite{GrKo+21}. Most notably, Nahar and others have investigated the intersection of software systems and ML models focusing on collaboration between teams~\cite{nahar}. They found collaboration challenges among and within teams through miscommunications, a lack of documentation, non-valued engineering, and unclear processes.
Such studies provide a rich, yet incomplete picture on the social and organizational aspects of building ML-enabled software. 
\begin{figure}[ht]
    \centering
    \includegraphics[width=\columnwidth]{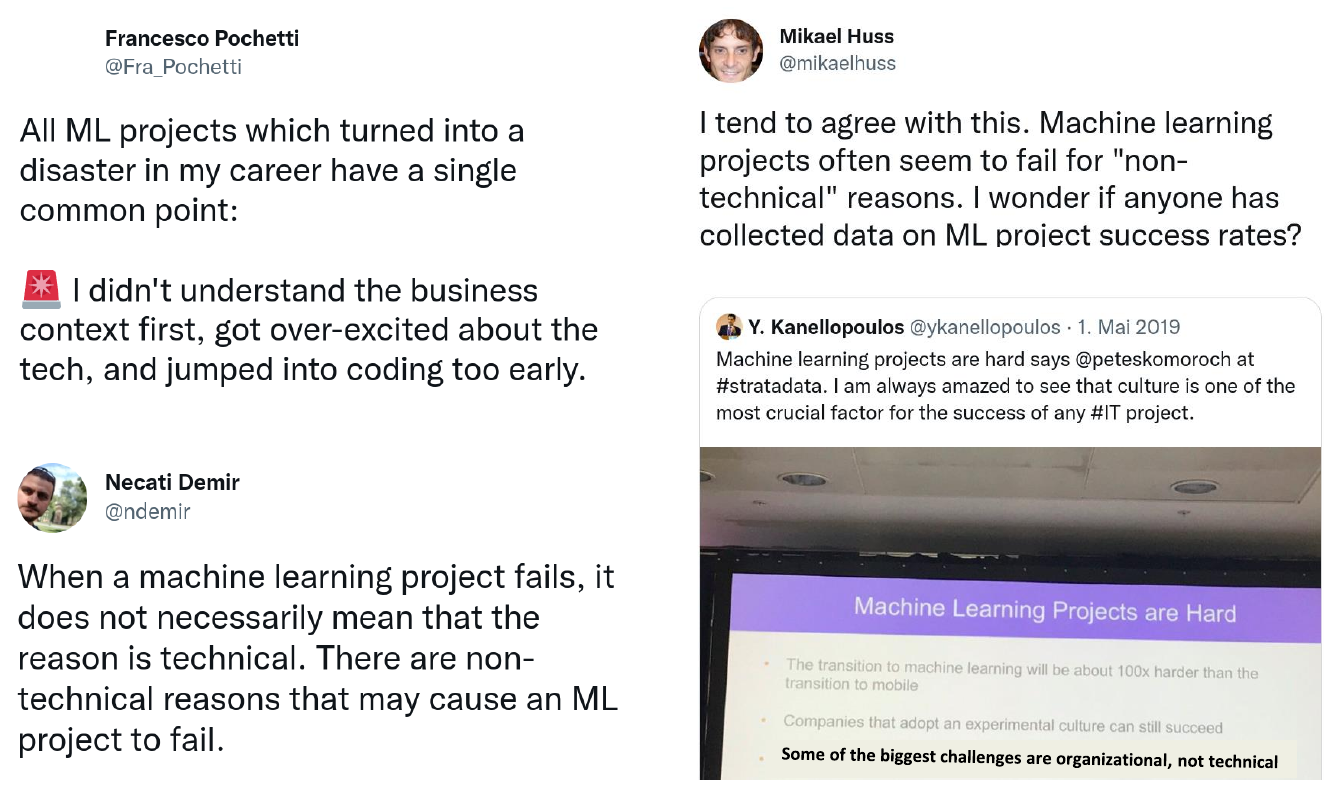}
    \caption{Anecdotal evidence of non-technical issues of failed ML projects.}
    \label{fig:tweet}
\end{figure}
Anecdotal evidence from blog posts, tweets, and (video) podcasts, such as in Figure~\ref{fig:tweet} hint that not solely tools and technologies are main factors for failed projects, but management, organization, and social aspects. Further, so called \textit{gray literature} has been shown to contain valuable insights to enrich literature reviews~\cite{Garousi2020}.
It is, thus, surprising that there is no systematic study that lifts this huge body of knowledge provided by practitioners who encounter these problems in the first place and find their insights valuable to share and discuss with each other.

We set out to explore what experts and leaders report on socio-technical challenges when building ML-enabled software. Specifically, we state our research question around the organization and management of teams and persons: \textit{What socio-technical challenges and what kind of solutions do practitioners discuss in regard to productionizing machine learning?}

To get unbiased opinions and discussions around the development of ML-enabled systems for answering our research question, we searched for an interest group whose members are focusing on ML-system development in a holistic way, that is, they are not focused on specific stages of the development process, represent a wide diversity of members in terms of roles, company size, and country of residence, and have a culture of sharing their activities in a suitable format. The biggest interest group that fulfills all of these criteria is the MLOps.community\footnote{\url{https://mlops.community/}}. A community with more than 11,000 professional members dedicated to share experiences when putting ML into production. 

We employ a research methodology known as {\em reflexive thematic analysis}~\cite{braunclarke, Braun2019} to identify thematic patterns and interpret them. We manually analyzed in a qualitative empirical study 73 videos published by the MLOps community, thereby analyzing more than 66 hours of video material\footnote{This roughly maps to 1,200 pages of automatic generated transcripts.}, making it the largest qualitative study we are aware of (cf.~\cite{CZF21, STV18, SSZ21}). This enormous corpus of data gives us insights from a large variety of companies with different sizes and diverse business goals. Since the practitioners originate from five different continents, we obtain a broad picture of socio-technical challenges, involving different cultures.

Our main contributions are the following: We found 17 anti-patterns whose causes are mainly of non-technical nature. That is, the reported anti-patterns of practitioners often originate from a lack of organizational decisions and guidelines, and undefined organizational processes as well as miscommunication between management and the development team. Specifically, we could identify three main areas in the organization of ML projects causing problems: leadership vacuum at management level, organizational silos, and communication within an organization. Practitioners mention that tools and technologies may mitigate the symptoms of anti-patterns, but the causes lie too often within the organization itself. This draws parallels to similar observations in traditional software development, such as Conway's Law~\cite{conway,conway2}. 

As second contribution, we contextualize previous research by discussing how challenges found before~\cite{nahar} can be traced to our identified anti-patterns and their causes. Thereby, we not only produce novel anti-patterns, but also confirm from another data source existing ones, strengthening the generalizability and trustworthiness of our and prior work in this area. We could enrich former findings due to the unique opportunity of evaluating interviews of mostly leaders (CEOs and managers) instead of developers. This novel perspective from a management point of view is a missing piece of a puzzle to have a holistic view on the development activities and challenges in building ML-enabled software. Furthermore, confirming and extending on former findings shows how practitioner communities can be leveraged to gain insights from samples with special characteristics and size that are otherwise barely obtainable.

Finally, we report back possible best practices to address the mentioned causes. Overall, we found substantial evidence that socio-technical problems are often the root cause of many issues that practitioners and research alike aim to address with technological solutions. By framing these problems as anti-patterns and rooting them at the social and organizational level, we raise awareness of this underrepresented line of research.

\section{Methodology}

To answer our research question, we conducted a qualitative empirical study. We investigated 73 out of a total of 210 videos published by the MLOps community using reflexive thematic analysis (RTA)~\cite{braunclarke, Braun2019, braun2019reflecting} to identify and analyse thematic patterns. RTA is a fully qualitative method from the family of thematic analysis and is widely used in psychological research, but becomes increasingly adopted in other fields, such as human computer interaction~\cite{brown13hci}.
In contrast to other methods for thematic analysis, such as coding reliability approaches, RTA allows researchers to engage with the data in an analytical manner rather than just providing summaries of what was said \cite{braun2021can}. This was necessary in our case, since we could not rely on a structured questionnaire that would prompt practitioners to speak about certain content (as in an interview study).  

 A key feature of RTA is `late' theme development, meaning that themes are generated from codes analytically. Themes represent shared structures, underpinned by a key concept~\cite{braunclarke, braun2021can}. Braun \& Clarke outline 6 steps in the analytical process:  familiarization with the data (1), coding (2), initial theme development (3), developing, reviewing and refining themes (4), defining and naming themes (5), and writing (6), which are run through recursively. This way, RTA enables to systematically analyze diverse and inconsistent data, such as conversations and, thus, gives larger flexibility to this kind of data compared to, for example, grounded theory and related methodologies~\cite{braun2021can}. Since we use an existing corpus of unstructured interviews and talks, this research methodology is ideal for such a use case. 
 \begin{figure}[htbp]
\centerline{\includegraphics[width=0.99\linewidth]{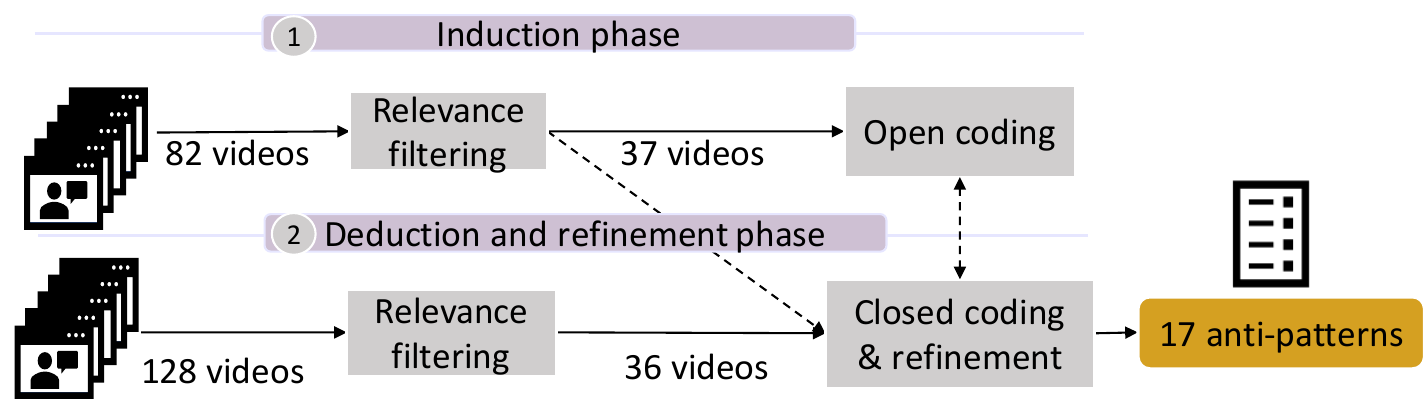}}
\caption{Two-phased qualitative research design.}
\label{fig:method}
\end{figure}

We apply RTA by having an inductive (bottom-up) phase and a deductive (top-down) phase as displayed in Figure~\ref{fig:method}. We provide the interim selection of codes that has been used (and refined) during the deductive phase and the later structure of themes as supplementary material~\footnote{\url{https://doi.org/10.5281/zenodo.7520777}}. Note that codes and themes evolve and change during the interpretation and iteration (one clearly does not start with the final themes). The interim codes represent the stage of our research when we identified that each theme consists of multiple symptoms (later evolved to anti-patterns), causes, and solutions (later evolved to recommendations).
Next, we explain our source material, demographics, and the phases of our methodology in detail.

\subsection{The MLOps community}
The MLOps community is an interest group founded in 2020 that is centered around knowledge sharing. Its major output formats are Meetups and Coffee Sessions, both held online and streamed via YouTube. 
Although the community's notion centers around pipelines and MLOps, the channels and interviews cover the whole development process, which makes them an ideal base for our analysis.

\subsection{Induction phase}
From the full set of 210 videos, we took 82 from the `Full Talks' playlist on the respective YouTube channel. Due to the long playtime of the videos and the relative little information contained in the video description, we conducted an initial filtering. For each video, we searched for the keyword `team' in the automatically generated transcripts and manually evaluated whether the speaker is speaking about socio-technical challenges. Although this singular keyword may seem too limited at first, we argue that socio-technical challenges arise around and within teams. Other non-technical aspects, such as ethics or legal concerns, that also occur in the development process are out of scope of our analysis. After the filtering, we yielded 37 videos from which we expected to contain at least one part relevant to our research question. 

Following the guide of Braun \& Clarke~\cite{braunclarke}, we familiarized ourselves with the video material while simultaneously correcting the automatically generated transcript (1). We used the identified relevant sections of the videos as starting points and transcribed all relevant content before and after to fully understand the context. We then started to develop initial codes (2) and already started to generate initial themes (3). In regular confirmation and validation meetings, we entered phase 4, in which we reviewed and refined our initial themes. Here, we reflected on differences and similarities in our perspectives, resulting in an iterative refinement of the themes. We already started with defining and naming themes (5) during this induction phase, which helped identify that structuring the initial themes into the presented form of anti-patterns/causes and recommendations is meaningful. For further iterations, we used the form of symptoms, causes, and solutions in our code collection. The rationale was that the themes we identified so far, had more than one challenge included but not necessarily all speakers mentioned each challenge. Some of them just described issues that we identify as causes. The initial themes developed to be the seven contexts and development activities presented in this paper.


\subsection{Deduction and refinement phase} 
To further engage with the remaining videos, we diverged from the original proposal of Braun \& Clark and used our extracted codes to identify relevant videos in the remaining 128 videos. We therefore read title and description of the videos as well as their topic tags, which are provided by the uploader for some recordings to label different sections of the videos. Another 38 videos were marked as relevant and we started again familiarizing with the data by correcting the automatic transcriptions (1). This way, we reduced the threat to bias our selection against our initial keyword. We then used our initial codes and themes to find similar passages within the videos. The goal of this phase was enriching and refining the themes with the additional data. Although we run through earlier steps (2-3), we focused mainly on refining (4), defining and naming (5), and writing (6).  We finally arrived at our structure of anti-patterns and causes. We reviewed all relevant passages again and extracted practitioners' recommendations. 

\begin{table}[htbp]
\addtolength{\tabcolsep}{-3pt}
\caption{Tracing of reported anti-patterns in the individual contexts to the corresponding sources.}
\begin{center}
\scriptsize
\rowcolors{1}{white}{rowcol!70!white}
\begin{tabularx}{\linewidth}{p{3cm}XX}

\hline
Context & Meetups [M] & Coffee Sessions [CS] \\
\hline
Model to product integration &3, 10, 17, 21, 35, 36, 43, 46, 47, 54,  56, 57, 64, 75, 84, 92, 94, 95.1& 20, 34, 35, 43, 44, 62, 69, 73, 90\\
Data producer to consumer integration&10,44, 46, 59, 64, 94 & 19, 26, 38, 39, 65, 89\\
Redundant development & 22, 29, 49, 68 & 13, 23, 26, 44, 49, 67\\
Headless-chicken-hiring & 10, 37, 46, 49& 19, 23, 25, 27, 38, 40, 44, 81, 89, 95\\
Résumé-driven-development & 5, 30 & 29, 38, 85, 86 \\
Hype-driven product creation &5, 10, 11, 30, 37, 45, 46, 49, 54, 55, 59, 60, 61, 62, 68, 69& 18, 21, 23, 35, 38, 39, 43, 44, 46, 56, 67, 75, 85, 87, 88, 89, 93, 94\\
Management vs. Data Science &29, 30, 58, 64, 65& 6, 21\\
\hline
\end{tabularx}
\label{tab:sources}
\end{center}
\end{table}

We maintained the tracing from a context or development activity to the according videos (cf. Table~\ref{tab:sources}). This allows us to mark a theme as `thick' (i.e., often discussed) or `thin' (i.e., rarely discussed). However, we refrain from a more granular mapping to individual causes or anti-patterns because interviewees reported patterns and causes in different levels of detail. Some only enumerate problems whereas others provide detailed discussions. So, while a quantitative weighing of a theme's importance may provide additional insights, we are careful to not lead the reader into misinterpreting frequencies with severity of challenges. Hence, we conservatively follow the fully qualitative approach of RTA and refrain from providing more detailed numbers.

\subsection{Demographics}
The final sample consists of 73 videos with a total of 66 hours of conversations and talks with a mean number of views being 628. The oldest video was recorded in April 2020 and the most recent video has been recorded in May 2022 and represents the end of our data collection. For each video of the original 210 videos, we used the YouTube-Transcription-API to derive automatic transcripts for an initial filtering step. 

\begin{table}[htbp]
\addtolength{\tabcolsep}{-3pt}
\caption{Demographics}
\begin{center}
\scriptsize
\begin{tabularx}{\linewidth}{lX}
\hline
\multicolumn{2}{c}{\textbf{Company characteristics}} \\
\hline
Location &  North America (50.\%), Europe (39.2\%), South America (6.8\%), Asia (1.4\%), Australia (2.7\%)  \\
Size & Large (40.3\%), mid-size (16.9\%), small or startup (35.1\%), other (7.8\%)\\
Sector &  Technology (31.1\%), consultancy (14.9\%), ML-platform (14.9\%), e-commerce (9.5\%), banking \& fintech (8.1\%), venture capital (6.8\%), other (13.5\%)\\
\hline
\multicolumn{2}{c}{\textbf{Speaker characteristics}} \\
\hline
Roles & Management (director, head, chief, CTO, CEO, founder) (31.3\%), leader (+ principal \& resident) (27.6\%), data scientists \& engineers (+ data-, ML-, MLOps-engineers) (22.7\%), product management (7.6\%), other (10.8\%)\\
\hline

\end{tabularx}
\label{tab:sample}
\end{center}
\end{table}

The speakers in the videos come from different, mainly US-American and European countries and mostly hold leading positions. Often, guests do not solely talk about their current position, but also report stories from former employees, colleagues, and discussions in the Slack community channels. That is, we do not necessarily see them as sole representatives for their respective companies, but rather as individuals experiencing the move from traditional software to ML-enabled software.

\subsection{Threats to validity}
Qualitative studies are naturally threatened by a subjective analysis process. That is, we could have focused too heavily on specific topics while missing others. This includes the manual encoding and generation of themes. RTA emphasizes that themes are an interpretation of codes and should not be seen as independent from the researchers~\cite{braunclarke, braun2021can, braun2019reflecting}. We held regular meetings in which we focused on building a rich and common understanding of the data and reflected on our process and assumptions. Further, we put our results into perspective with prior work, thereby validating the anti-patterns and causes found in our study. 

The initial selection of the keyword and relying only on a single community may pose an additional threat to external validity. With its over 11,000 members, spanning several continents, we are convinced that this represents one of the most generalizable sources available, especially when compared to singular company papers. Moreover, by having a two step process, we extend our initial filtering process with topics mentioned around teams, such that further socio-technical challenges appear. Our surprisingly high alignment with findings of other papers that use structured interviews provide further evidence of the validity of our findings.

\subsection{Results Overview}

We group our findings into three main areas: organizational silos, communication within an organization, and organizational leadership vacuum. In these areas, we found 7 contexts or development activities in which, in total, 17 anti-patterns can emerge. The next sections are structured as follows: We first describe the context of reported anti-patterns. Then, we list all corresponding anti-patterns (denoted with \textcolor{patterncolor}{AP}) together with citations that are exemplary for our findings. Afterwards, we describe their causes (denoted with \textcolor{causecolor}{C}) and suggested recommendations (denoted with \textcolor{recommendcolor}{R}). Finally, we put our results into perspective with prior work in Section~\ref{sec:discussion}, thereby, confirming earlier findings and validating ours.

\section{Organizational Silos}
Organization silos often manifest when multiple teams work on disjoint tasks, challenging data and model transfer. 

\pattern{Model to production integration}
\label{ref:mvp}

\quotebox{How do I hand off this model to this other team? And there's a lot that goes into it, it's the actual version of the code to instantiate the model, it's the right set of weights. You probably want some way to track metadata associated with it, in case there are scores to be able to interact programmatically. And so at its core, this suddenly goes from what sounds like it should be a pretty easy problem to 'I'm spending weeks and meetings and emails trying to physically figure out and help this person instantiate my model and load my weights.' [..]
And then, it gets worse because you try to build a new version of the model and instead of that person being able to grab the new version, they have to go through this whole stupid cycle again.}{M36}

Many times practitioners describe the process of getting an already developed model into a production stage as especially tedious and error-prone. This problem was reported to occur across all stages of company maturity. A model is developed by either a single data scientist (integrated team, as defined by Nahar and others~\cite{nahar}) or a dedicated data science team (center-of-competence). However, the developing party is always dependant on another organizational entity (e.g. another team member or another team) to integrate and deploy said model (model handover). 
Furthermore, the model handover is not only a handover of the model as such, but oftentimes includes feature pipelines that have to be implemented in ways that differ substantially between development and production environment. 
The interdependence between the two organizational entities increases when the code bases (i.e. model development vs. production) differ in terms of languages and tools. This can lead to the necessity to reimplement whole pipelines, which then have to be validated again by another person or team. The following anti-patterns have been observed independent of whether the model team or an engineering team was responsible for productionizing the model.

\symptom{(AP1) Long release cycles} Symptomatic outcomes are long release cycles, back and forth between model developing and product team or even the complete inability to productionize existing models. The two teams are tightly coupled and spend a lot of time discussing all necessary details. While direct communication is often times preferable, it can become an obstacle if there are no code abstractions, documentations, and clear processes defined so that the teams get together for every detail in every new version. In some cases, the whole model has to be rewritten by the product team, leading to long intermittent phases between two versions as well as technical challenges such as assuring that the rewritten model holds the same properties as the original one. 

\symptom{(AP2) Tension between teams} The collaboration between the model developing entity (often data scientists) and the entity that is responsible for putting it into production (often software engineers) is characterized by reciprocal complaints that are usually driven by a lack of understanding. The speaker in M3 gives a brief example of what communication between the two entities looks like in practice: \simplequote{So, you had these multiple jumps and it was a daily exercise of complaining about the other person. So the software engineer would complain to the data scientist saying `I don't understand what's going on here. It's crazy, it's too complicated' and the data scientist is going to the software engineer `It's not complicated, it's dead easy, how can you not understand it?'}{M3}.

\symptom{(AP3) Blocked data scientists} Long release cycles and the tight coupling of the two teams implies that data scientists are not able to focus on developing or improving new models, but invest their time in working on the production side.  The same is the case when the data science team is supposed to productionize the model themselves: While model development is often seen as the core task of a data scientist, they are now engaged with the production part and are therefore no longer developing or optimizing new models.

\cause{(C1) Clash of cultures and tools} 
Speakers see the causes for painful handovers in the inherently  different cultures of software engineering and data science, where both have their own values and expectations. For example, model development is usually research-oriented, leading to a reported lack of product-centered thinking in data science. Artifacts, produced in the model development phase, are often not tested and do not follow code conventions. Moreover, both fields use different tools, languages, and frameworks, leading to 
different terms and concepts, which further enforces communication issues. \simplequote{One of the challenges I have with data scientists is to understand the kind of language they use [..]}{M3}

\cause{(C2) Missing engineering competence} 
Data scientists and software engineers alike lack skills to sufficiently understand the other side. 
Software engineering practices, such as code quality, design patterns, automation pipelines, virtualization, and scalable infrastructure are often unknown to data scientists. On the other hand, the software team operationalizing the model is missing key data science skills and is thus treating the model as a black box, which, in turn, necessitates check backs from the engineering team; all of this leading to production delays.

\recommendation{(R1) Model registry and feature store} A common approach is easing the symptoms of anti-patterns instead of targeting the causes. That is, technological solutions can circumvent some of the issues existing when integrating a model to production. 
Especially, model registry and feature stores enable a cleaner interface between organizational entities and a clear contract in terms of documentation and standards between them. Model registries are repositories in which the model artifacts are stored together with all relevant information (e.g. code, data, parameters, metrics). Feature stores, on the other hand, enable versioning of feature calculation for training and production in a central database, eliminating mistakes and long validation cycles. By using these tools, the organizational entities are forced to work in systematic and structured ways. This is also possible without these tools, but requires non-technical solutions, such as guidelines and contracts.

\recommendation{(R2) Restructure teams to be cross-functional} While technological solutions focus on bridging existing gaps, many speakers mention that an organizational restructuring from functional teams towards cross-functional or integrated teams have helped them to bridge the gap between data scientists and software engineers. When working closely together, people are likely to share skills and develop a common understanding of the product they are developing. 

\recommendation{(R3) Pair data science and software engineers} Especially for helping data scientists to write code of high quality that adheres to the standards of the software engineer (i.e., “production code”), pairing and code reviews are two non-technical solutions mentioned. While this can be performed after a restructuring within each of the teams, the speaker in M58 advises to also perform this kind of exchange between functional teams: \simplequote{[..] a great way for us to upskill the data science team is to pair with them work on a project [..] and show them how to do that correctly and why it's important. I've never seen much success with a hostile engagement like blaming, finger-pointing. It's not productive.}{M58}

\recommendation{(R4) Translation work between the different roles and common understanding of the goal} Different fields and values lead to different vocabularies used by their representatives. That is why practitioners recommend addressing this by either building a common dictionary with terms and a corresponding definition or introducing roles that have an understanding of both fields and act as a mediator between them. Having a common language across different roles and teams enables understanding of products as shared responsibility and fosters empathy for the individual challenges each of the fields has.

\pattern{Data producer to consumer integration}
\quotebox{So, the backend handing over that data object, that has been a very painful process and primarily not so much because of technological restrictions, but much more because organizationally the data team has been quite separate from the backend team. Trying to get stories into their system has been quite difficult. And I think that's something that I still see today: where it's very hard to convince other teams to expose some kind of data that you would need or to get them to fetch some data that's not directly relevant for them or not relating to some kind of use case.}{M46}

\normalsize
Larger companies often have multiple departments and are organized in ways to treat them as distinct business units. Hence, there are organizational units that produce data (e.g., monitoring industrial processes, gathering customer data) and store them in their own backend systems. Especially ML teams require to consume that data to build new models. The separation of the two parties result in several anti-patterns.

\symptom{(AP4) Requesting data is hard} Symptomatic for this pattern is the restricted access to data for the consumer by the producer. Practitioners describe that it is hard to convince the producing organizational entity to expose data or fetch specific data points that would be necessary for the consumer's use case. Oftentimes, the data producing team is not directly involved in these use cases and therefore fetching, exposing, or fixing data points are constantly weighed against the use case of the producing team. Hence, requests are often shut down or unanswered since providing data often raises costs for the producer without any benefit.

\symptom{(AP5) Tight technical coupling between consumer and producer team} An unclear handover process of data leads to close technological entanglement between teams. This results in code and data dependencies across organizational units. \simplequote{Feature calculation against operational databases, this is a really horrible one, that I never want to repeat in my life again. Both from computational points of view but also from just separating out contracts.}{M46}. 
Further, implicit assumptions by the producer team can lead to silent changes of definitions or calculation of certain data points, which is not immediately visible in the outcome on the consumer's side. These silent changes can  lead to false predictions of data services further down the line and make debugging really hard. 

\symptom{(AP6) Partial data availability} Incomplete data available to the data science team may result in differences between training and production. If this is unknown, a deployed model may likely suffer a degraded performance as the distribution of training data does not reflect the data distribution in production.

\simplequote{He was creating a product, and when he was doing his training data, everything was fine. But then when it went live, this machine learning algorithm wasn't getting the full story all of a sudden. Later he found out just by coincidence that his training data was just one stream of the whole picture. So, it was like `Oh wait a minute! That's why everything's going haywire when I put it out live!'.}{CS19}

\cause{(C3) Lack of awareness and common goals} Practitioners mention that the organizational division between consumers and producers, where each one is following their own use case, is a major cause for the problems in collaboration between the two entities. The producer lacks awareness and concern about the data and does not see the value in spending time and effort on fetching and maintaining certain data points. Further, the consumer's requests can be in conflict with the producer's own product and success metrics. 
It is even possible that the producing part is not aware of any downstream dependencies of their data output which enforces silent failures: \simplequote{How do you give them [the producer] the right feedback loop, so that they understand that if they change something it's potentially a breaking change for a different unit somewhere else, because they [the consumer] use that data as an input data for some of their machine learning systems.}{CS89}

\cause{(C4) Missing documentation and unclear responsibilities}  Especially in larger organizations with a centralized data structure (e.g. a data lake or a data warehouse), it can become intransparent who produced the data and who is therefore responsible for maintaining it. Furthermore, definitions might lack the necessary clarity, which leads consumers to either not use the data or produce results which can not be trusted. Such a lack of proper documentation makes it hard for consumers to trust the data and exposes them to the risk of having misbehaving models during production where only by chance one can find the cause in the data sources.

\recommendation{(R5) Raise awareness of data producers} Changing the mindset of data producers towards understanding how important their data can be for other teams is a common recommendation. Some practitioners mention that this remains an unsolved challenge within their company; education and feedback are mentioned as possible paths. Education focuses on emphasizing the role each team plays within the company's wider scope and what kind of consumers exist, including their specific needs and risks. Further, a feedback mechanism that shows the utility of the data by the consumer and also enables the producers to see the influences of their changes directly. 

\recommendation{(R6) Central platform as a contract between producer and consumer} A common recommendation was to use a central repository or platform to allow for discovering and accessing of data. Still, missing documentation and trust in the centralized data as well as the high level of structural independence needs to be addressed. Therefore, speakers suggest that the platform itself should act as a contract, including quality metrics and documentation of the ingested data.

\section{Communication}
Missing communication has been reported to be a major problem for causing a redundant development of features, tools, and infrastructure in organizations as well as for a source of tension between the data science team and the management.

\pattern{Redundant development}

\quotebox{Let's say you have two models which are quite similar. Both use the same feature. Both are developed in different teams. And then, those teams start to develop exactly the same feature in slightly different ways but semantically with the same meaning.}{M29}

Companies with multiple teams and multiple use cases frequently report the situation that similar features (i.e., preprocessed raw data acting as inputs to the ML model) are subsumed by multiple models. Ideally, those features should be defined and preprocessed only once, but, in practice, are redundantly developed, wasting time and resources. Even worse, this situation is not unique to features, but redundant infrastructure and tool development, such as pipelines, database systems, scaling infrastructure, has been reported to potentially occur in every company with multiple use cases. 

\symptom{(AP7) Features are not discoverable, accessible, or reusable} Different teams are developing semantically similar features, but the features can not be reused. Even if they are used in different teams, there is no direct connection. That is, when a feature is updated or the constraints on the calculation change, every team has to correct them, leading to repetitive work. 
However, it has also been reported that features are given similar names, but have dissimilar definitions. So, the interpretation of domain concepts that may result in ML features is subjective to the team's perspective and use case.

\symptom{(AP8) Redundant ML infrastructure and tools} \simplequote{I'm sure the tradition is: I have a problem, I build a model. I want to keep updating my model. I'm going to build some infrastructure to update my model. Now, I'm an infrastructure team for my model and so is everyone else.}{CS23}. Building an ML model usually requires an infrastructure for obtaining and preprocessing raw data in a scalable way as well as conducting experiments and training runs. This leads to redundant infrastructure and the accompanied tools in the organization. 

\symptom{(AP9) Shadow IT} Since every team is developing their own infrastructure, there is a chance that the development happens although there is a central infrastructure provider who is normally responsible for hosting and maintaining infrastructure. The disconnect between the teams and the infrastructure provider leads to ungovernable self-hosted IT within the teams. Sometimes, this means that organizational security standards, such as protection of APIs, are ignored and, therefore, especially in the context of data, holds a major risk to privacy and security. 

\cause{(C5) Missing intersections between teams, documentation, and trust}  In this decentralized scenario, all teams work on different use cases independent from each other. The absence of a natural intersection between the teams leads to little or no communication, meaning that the teams do not know about each other's developments (e.g., features or tools). 
Even if the developments of a team are potentially discoverable by other teams, it does not imply their usability. The isolation of the teams and use cases does not set the required incentives to build and document in a reusable fashion. This leads to missing trust by potential (re-)users of the features or infrastructure: 
\simplequote{Can I actually reuse all those existing features that other team has developed? How do I get that trust and confidence that my use case can also be served? Right now many times, a team would be `Hey I need to get my use case out very soon. So, I don't really care about documenting well, putting the right description about what the features are'. So, everybody will be a bit lazy in making sure they have done a good enough job so that people can understand.}{CS26}.

\recommendation{(R7) Unification and discovery through centralization} Practitioners mention centralization as a key concept to unify different tools as well as enable discovery of features. The central management of tools and infrastructure takes the burden from individual teams and gives them more time to focus on their use cases as well as enables easier communication between teams. Feature stores are again mentioned to be key to a central feature management. While it enables not only a clean translation of features for training and production, it further enables reusability and shareability across organizational entities. Besides feature stores, any other concept that involves central management of features and corresponding managing infrastructure can help. 

\recommendation{(R8) Showcase meetings} For better understanding of each other's work and progress, some practitioners propose regular meetings in which a team presents its work, their use case, and their progress and learnings. This fosters understanding of use cases, the development of intersecting interests and infrastructures, and education of new technologies.

\pattern{Management vs. Data Science}

\quotebox{All of the higher-ups were just looking at the data science team as a money suck. All they were doing was walking around. They weren't putting anything into production. They weren't doing much: looking at data, asking for more data. And, they weren't actually producing anything. And so the C-level executives: ‘Hey what are we paying these guys for? What is going on here?’ And they couldn't figure out the value.}{M29}%

Naturally, the management needs to be informed about the progress and state of development, much like in traditional software development. Tensions between management and development are, among others, one reason for introducing self-organizing teams and agile principles. However, the progress and outcome in ML development is inherently uncertain, which further amplifies communication hurdles. 

\symptom{(AP10) Data scientists struggle in their roles and get burned out} Data scientist struggle in their roles, feeling like they are not delivering enough value and are not producing tangible results for the company. Practitioners link this to feelings of burnout due to never ending iterations that finally do not lead to valuable results. Without documentation, practitioners find it hard to look back on their own work and see their contribution.

\symptom{(AP11) Tension between management and data science} In the analyzed videos, data scientists as well as management report that communication can be difficult with each other. While management expects the data scientists to deliver value to the company, data scientists follow their research-oriented workflow. This increases tensions as the open research workflow may not be aligned directly with the business value. Usually, metrics do not map directly and the effects on the ML model are not directly visible. 
The frustration managers report goes even further into the accusation that data scientists neither want to be managed nor want to be confronted with their own value in the company: \simplequote{When you talk about data scientists: if I use the letters ROI, I get hate mails. And I've done that before! I've posted: `You have to have ROI, you have to be connected to business problems.' I get hate mails. Data scientists hate it. [...] A lot of data scientists come straight out of academia and it's a different paradigm.}{M30}

\cause{(C6) Missing data science process} 
A stated cause for the anti-patterns reported above are missing standard development processes, such as SCRUM that the management knows and can align with its expectations about the development progress. The research-driven approach of model building contains too many uncertainties: 
\simplequote{Software is very results driven, where it's very easy to be like: `Hey, I build this piece of code and it should be x+x equals y. [..]' The input and the output should be very clear. Sometimes data science isn't that way [..] where it's more research-based and kind of long-term projects, there's a likelihood of failure.}{CS21}. The combination of inexperienced management that does not fully understand the development process of data science projects and the absence of standard development frameworks for data science lead to unstructured approaches that are often not accompanied by documentation and reporting. 

\cause{(C7) Communication mismatch between technical and non-technical people} The communication mismatch between technical and non-technical people is seen as a major cause for diverse problems not only for data science but also for software development. The inability of technical people explaining their problems and achievements in a way that is understandable and relevant for the business is the root of tensions between the two groups. \simplequote{I’m like `Hey, check out this cool RMSE graph!' And with almost no labels on it and show it to a marketing team. And, they look at it as if they’re meeting an alien for the first time. They’re like `what is this?'.}{M58}

\recommendation{(R9) Strong processes} Building strong development processes for data science in general is  a key recommendation from practitioners. They additionally recommend defined processes on how to intersect and communicate to management and other business-oriented entities. Tailoring the agile framework to the experiment-driven building process of ML models might be a way, but research is missing on this part.

\recommendation{(R10) Documentation and reporting} Documentation has been recommended as a main tool to communication with the management and to enable retrospective analysis of the progress made so far by the data scientists themselves. Documentation is seen as the foundation for writing reports and preparing presentations, in which data scientists should focus more on communicating in terms of business goals than specific technical metrics. This is an interesting insight as this, at least partially, diverges from the low documentation approach of agile practices in software development. \simplequote{Having that
ability to always look back, there’s a lot of value there. And a lot of people just are `Oh, I just analyze this data and write up a powerpoint and then be done with it'.}{CS21}

\section{Leadership Vacuum}
We identified that many stated problems are due to a leadership vacuum. That is, there is no authority that either acts at all or with sufficient knowledge about data science. 
\pattern{Headless-Chicken-Hiring}

\quotebox{Everyone kept thinking that the solution would be to hire more data scientists. So, for me it was going back to leadership and the head of engineering and saying: 'No, please stop! You can add [..] as many as you want, you're not going to solve the major issue that we have.'. [..] And this title conundrum came up with every single one of them. It was continuously 'Well, we can just hire a data scientist. Because we need ETL pipelines and we need somebody to make sure that they architect a very good relational database and then push all that reporting out to our multiple users', and I was like: '[..] Is that the right person that you should be hiring?' And they're like: 'It has data in the title'.}{M37}

Putting more people in a team, especially late before deadlines, has already been debunked for traditional software engineering~\cite{brooks}. This ineffective management decision becomes even worse when the wrong people are hired, which have not the skills that are actually needed. Uninformed and non-fitting hiring has been reported by multiple practitioners. This situations seems to be more common in start-ups or companies that are just starting to build their first data-science use cases.

\symptom{(AP12) Staff with insufficient skills} 
The staff that has been hired is not equipped for the task at hand. An reported example is that people with a strong machine learning background are hired, but are supposed to build and maintain pipelines or applications where more engineering-skilled staff would be required instead:
\simplequote{The larger question that came up at that time [when they needed more complex pipelines], was `Well, the data science team will do it'. And I was like: `We have a lot of statisticians. We have a lot of people that are really good with machine learning algorithms. We have people that [..] could build out the pipeline. But they don't have experience in maintaining it. They don't have the [..] the software engineering background [..]'.}{M37}. 

\symptom{(AP13) No product for data scientists} 
A closely tied anti-pattern may rise when after hiring a data scientist for a specific task, there exists no further use case for that person. Even worse, if the initial use case turns out to be solvable without her expertise, it can have a negative impact on social structures and productivity between people and teams within the organization: \simplequote{After that [the initial project], the person may be pushed to create something. Or, in other scenarios, be considered the person that's not doing much work because the company and the product team hasn't found another product that they need a data scientist on.}{M37}.

\cause{(C8) Unclear roles and titles}
Multiple speakers mention that roles and titles in the field of data science and machine learning are consistently undefined. That is, a certain job title can imply different tasks and required skills depending on the company advertising the job. This complicates the hiring process for management and leads them to often simply search for data scientists while not accounting for the wide variety of skills people may have. Furthermore, successively more diverse job titles appear: \simplequote{
We have this ML scientist, we have the amorphously defined ML engineer. Then, we bring in software engineering machine learning. We have site reliability engineers. We have all these different roles. Now, MLOps is becoming a thing. I wouldn't be surprised if we start to see MLOps engineers.}{CS23}.

\cause{(C9) Uneducated hiring} Due to the complexity that comes from undefined titles and job descriptions, hiring managers require significantly more knowledge. Since this field is relatively new and in constant change, it seems hard for leaders to keep up with the demands and skills. Moreover, a missing communication about the actually required skills within the team seem to be a further cause for this. The result is that data scientists represent a fall-back solution for hiring.

\cause{(C10) Skill shortage on the market} The fast-growing industry around artificial intelligence has risen the demand for skilled people on the job market. This amplifies hiring challenges and may lead to hiring people with wrong skills due to the absence of people with the required skill set.

  \recommendation{(R11) Hire for the skills and potential rather than roles} 
  A clear definition of the task, people are hired for, and the relevant skills needed to solve that task are seen as key foundations of successful hiring. Especially when titles are ambiguous, hiring for specific skills bypasses drifts of expectations in applicants. Speakers mention that the actual skills needed are more related to engineering rather than traditional data science. Recommendations go so far that, in situations where it is yet unclear how much data science will be necessary, hiring a more engineering focused person with the ability to build up data science skills is more future proof. Moreover, if the future development of a product is unclear, it may be more suitable to hire a data scientist as a contractor or consultant.

\pattern{Résumé-driven-development}

\quotebox{One of the developers, this guy, is very, very good in Scala. But, the problem is that the rest of the team used to work only [in] Python [..] This guy made the whole data processing pipeline in Scala and just delivered the mashed data for data scientists. And basically this guy [..] put in his résumé or his LinkedIn something like ‘Yeah, I used to do tons of engineering in Scala’. But 
[..] the whole maintenance took days [..] 
or most of the time the code itself was not manageable at all.}{M5}

Developing and productionizing an ML model involves several stages at which decisions are made about which tools and technologies are used. Ideally, decisions are driven by the business goal, the available resources in terms of infrastructure and developer skills, and organization aspects. Résumé-driven-development is a phenomenon that just recently was defined as an interaction between recruiters and software professionals in which knowledge and experience with new and trending technologies is overemphasized~\cite{fritzsch21}. Practitioners in the MLOps community mention this to further influence decision processes within teams.

\symptom{(AP14) Developed tools and models do not match the team or product goals} 
The decisions process about the choice of developed tools and libraries is strongly influenced by individuals. If there is no authority and a limited experience in the team, decisions may be driven by personal benefits, and not dedicated by the ML use case. If the responsible individual leaves the company, the project gets stale and is not maintainable by the remaining team members.

\cause{(C11) Data scientists do not identify with the business value} A reason for this anti-pattern is the disconnection of tech-focused people from their outcome inside the company: 
\simplequote{They [the developers] are not allowed to participate in the actual impact of the thing that we're doing. [..] So, I'm going to choose the next architecture and then it gets centered around those kinds of things [..]}{CS29}.

\cause{(C12) Missing decision maker} When it is unclear who will finally decide or how the team is going to decide which technology stack will be the most suitable, it is possible that everyone within the team is having a strong opinion on the tools. This chaos enables résumé-driven-development.

\recommendation{(R12) Rely on organizational knowledge} When deciding for the tools and technologies, it has been suggested to rely on existing technologies in other teams. This can act as a fallback guideline that triggers when there is no authority available or uncertainty about which technology to employ.

\pattern{Hype-driven product creation}

\quotebox{Monetizing data is different than just putting something into production slamming machine learning in. [..] 
No, you really shouldn't do everything [with ML]. Machine learning doesn't do! Stop! You can do a lot of what you want to do using traditional development methodologies or basic analytics.}{M30}

Identifying a use case that benefits from ML is challenging and should not be driven by hypes to generate profit. The anti-patterns emphasize the importance of proper requirements engineering, including suitable product metrics.

\symptom{(AP15) Stuck with proof of concepts} The organization keeps making proof of concepts (PoC) for algorithms but most of these never make it into an actual product. The speaker in M62 calls this phenomenon “proof-of-concept-hell”, where even successful PoCs never make it into production. The reason is that the product itself is not relevant for the user or not presented in a proper way. 
Sometimes, data scientists get stuck in endless iterations and optimizations of a single metric (e.g. accuracy). Ultimately, this process is not tied to the actual business value: \simplequote{You're really excited about technology. You just want to build something in one direction. It's like: `Okay, cool, how deep can we go? [..] How advanced can we make this model?'. It's really important [..] to really tie that back with exactly what you're saying: You want to understand what's the ultimate goal.}{M61}

\symptom{(AP16) Perception of ‘Everything can benefit from ML!’} Instead of a clear analysis of which aspects of a product can benefit from ML and the exploration of traditional development methods, there is a perception that everything should be done using ML. 
There is no clear roadmap or strategy of how ML will help the organization achieving a certain goal. This perception overrules fundamental business and project decisions, such as what is the value of an ML-enabled product feature for the customer, or how do we monetize ML.

\symptom{(AP17) Product is not feasible due to missing data or talent} Companies are very clear about what they want to do, but do not have the right data for their goal. Or, they cannot build it because they do not have the required staff. This seems to be a major problem of organizations that try to on-board a data product: \simplequote{
So, people will often think about what types of ML they want to be doing, but they won't have the correct type of infrastructure to get the data that they need [..]. So that's usually when people are rushing into something and then they realize: `I have zero training data'}{M61}.

\cause{(C13) Missing organizational strategy, governance, and structure}
Speakers report that organizations are missing a data strategy when introducing an ML product for the first time. Missing suitable data is a major reason, which further relates to unmanaged data collection processes. They identify this as a management or leadership problem, since this cause manifests already before the development of the ML-enabled software. \simplequote{For any company, of any size, when they're trying to create that new data product, the majority of the time, they really haven't collected the data they need. Because it's either changed too much, they don't have enough of it or the process for setting up that collection period wasn't done by somebody that wasn't [..] organized. And if you're not organized in data collection, you don't usually have a good end.}{M37}.  

\cause{(C14) Management lacks knowledge about ML} Practitioners report that the management has no clear understanding of what ML is and how it works. They further complain that this leads to underestimating the differences between a data-science product and a traditional software product. Often, the engineering effort put into productionizing a model and its influence on a return on investment is underestimated. This, in turn, creates the impression that every product can and should have an ML component. Simultaneously, companies seem to experience pressure to integrate ML into their products to stay competitive on the market, which enforces building products without the executive knowledge to manage them: \simplequote{From 2016 to 2019 maybe, there was a big emphasis with CTOs, CIOs, and CEOs that we had to do machine learning, we had to do data science, because if we weren't, competitors might be doing it, or I just won't have anything interesting to talk about with my other CTO friends when we meet for coffee.}{M45}

\cause{(C15) Missing translation between different stakeholders} 
Speakers emphasize how hard it is to find people that have both, the management and the technological skills, to successfully guide and develop machine learning products. This especially enforces the disconnection between development teams and stakeholders, which is mentioned as a key challenge that most speakers do not have a clear answer to. A result of the poor translation of the business or user problem into an ML question are metrics that do not reflect the actual problem and this ultimately leads to the wrong metrics being optimized. 

\cause{(C16) Missing production metrics}
Missing production metrics causes feasible PoCs to be unsuitable in production. If data scientists are not evaluating their models on metrics that are meaningful for the product (and customer), they can not judge whether a PoC is actually successful. 
\recommendation{(R13) Process to identify feasible use cases and product roadmap}
Recommendations intend to raise awareness that ML will bring a lot of complexity to the product.
So, most speakers recommend to rely on more traditional engineering methods and heuristics: \simplequote{If you can avoid using machine learning, then do it!}{CS56}.
Finding a reliable process to identify and evaluate use cases is crucial before starting to iterate on the problem itself. Therefore, the interviewees suggest multi-step processes that must pass before the product idea is given to the data science team: \simplequote{So, first you get sales, they go through sales. Then, you have a subject matter expert that looks at it, that knows a bit of data analysis. And then, if it passes through these filters, it goes through that AutoML to make sure that there's possibility on that end. And then, it can go to the data scientist.}{M60}. A further suggestion is to write a proposal that states feasibility estimates of the final product and data availability. It should be evaluated by multiple organizational units to result in a full roadmap that can be given to the data scientists implementing the product.

\recommendation{(R14) Understand customers and keep them close}
Practitioners recommend to keep problem owners and other stakeholders close to the data science team and start by understanding the actual requirements. 

\recommendation{(R15) Education} The management needs the ability to evaluate and identify use cases for ML. Education through workshops is recommended for emphasizing data availability. If a product is evaluated to lack data, setting up a governed and organized data collection process is required.

\begin{table}[h!t] 
\caption{Confirmed findings and corresponding anti-patterns (\textcolor{patterncolor}{AP}) and causes (\textcolor{causecolor}{C})}
\begin{center}
\addtolength{\tabcolsep}{-3pt}
\scriptsize
\begin{tabularx}{\linewidth}{XXp{0.6cm}}

\hline
 Challenge & Confirmed finding & Ours \\
\hline

\multicolumn{3}{c}{\textbf{Section 5 in Nahar and others~\cite{nahar}: Requirements and planning }} \\
\cline{1-3}
 \rowcolor{rowcol!70!white} Product requirements require input from the model team & Lack of ML knowledge in managers& \causebox{C14} \\
\rowcolor{rowcol!70!white} Model development with unclear model requirements is common & Model teams may receive some data and a goal to predict something with high accuracy, but no further context & \causebox{C16}\\
\rowcolor{rowcol!70!white} Provided model requirements rarely go beyond accuracy and data security & Rarely metrics beyond accuracy&\\
\rowcolor{rowcol!70!white} & Ignoring qualities such as latency or scalability leads to integration and operation problems & \\
ML uncertainty makes effort estimation difficult &	Science-like nature of data science makes it difficult to set expectations or contracts & \causebox{C6}\\

\cline{1-3}
\multicolumn{3}{c}{\textbf{Section 6 in Nahar and others~\cite{nahar}: Training data }}\\
\cline{1-3}
\rowcolor{rowcol!70!white}Provided and public data is often inadequate & Training data is often insufficient and incomplete & \patternbox{AP6}\\
Data understanding and access to domain experts is a bottleneck & Insufficient data documentation& \causebox{C4}\\
\rowcolor{rowcol!70!white}Need to handle evolving data&Data sources can suddenly change without announcement& \patternbox{AP5}\\
In-house priorities and security concerns often obstruct data access & Model teams have little negotiation power to request data & \patternbox{AP4}\\

\cline{1-3}
\multicolumn{3}{c}{\textbf{Section 7 in Nahar and others~\cite{nahar}: Product-model integration}}\\
\cline{1-3}
\rowcolor{rowcol!70!white} Team responsibilities often do not match capabilities and
preferences &	Hard to convince management to hire engineers & \causebox{C9}\\
Technical jargon challenges communication &	Differing terminology leads to misunderstandings & \patternbox{AP2}\\
 Code quality, documentation, and versioning expectations
differ widely and cause conflicts &	Data scientists do not follow the same development practices or quality standards & \causebox{C1} \causebox{C2}\\

\hline
 \multicolumn{3}{c}{\textbf{ Section 6 in Kim and others~\cite{kim2018}: Challenges data scientists face}} \\ \cline{1-3}
 \rowcolor{rowcol!70!white} Challenges related to data& Data availability and quality& \patternbox{AP4} \patternbox{AP17} \causebox{C4}\\
\rowcolor{rowcol!70!white} Challenges related to people & Convincing others data science is valuable and convincing them to collect data  &\\
\cline{1-3}

\hline
 \multicolumn{3}{c}{\textbf{ Section 7 in Kim and others~\cite{kim2018}: Best Practices to improve data science}} \\ \cline{1-3}
 \rowcolor{rowcol!70!white} Clarifying the goal of data science early in the project & data scientists can be tempted use specific techniques everywhere & \patternbox{AP16}\\
\cline{1-3}
 
 \hline
 \multicolumn{3}{c}{\textbf{Section 4 in Arpteg and others~\cite{arpteg2018} Organizational challenges }} \\ 
 \cline{1-3}
\rowcolor{rowcol!70!white} Effort estimation & In ML it is unclear if and with what effort a goal can be reached & \causebox{C6} \\
  \rowcolor{rowcol!70!white} Cultural differences & Data scientists and software engineers hold different values & \causebox{C1} \\
 & Tensions between product management and data science &\patternbox{AP11}\\

 \hline
 \multicolumn{3}{c}{\textbf{Section 4 in Granlund and others~\cite{GrKo+21}: Multi-organization problems }} \\ 
 \cline{1-3}
 \rowcolor{rowcol!70!white} Data integration problems & Data sets can not be moved, but all organizations need access &\patternbox{AP6}\\
 
  \hline
 \multicolumn{3}{c}{\textbf{Section 5 in Amershi and others~\cite{amershi}: Best practices }} \\ 
 \cline{1-3}
  \rowcolor{rowcol!70!white} Data availability, collection, cleaning and management & Availability and collection & \patternbox{AP6}\\

\end{tabularx}
\label{tab:findings}
\end{center}
\end{table}
\section{Discussion \& Conclusion}

\label{sec:discussion}

We have found a large range of anti-patterns and causes from different organizational viewpoints. Although our study is not confirmatory by design, we could reproduce several findings from related work~\cite{nahar, arpteg2018, kim2018, GrKo+21, amershi} and match them with ours. 
The result of this process is shown in Table~\ref{tab:findings}, where each heading indicates the related paper with its section and context, in which challenges have been reported. Column challenge  corresponds to a heading or paragraph within the corresponding section of related work, and the confirmed findings column refers to the exact finding we could reproduce.

Overall, we found in all papers overlap, demonstrating the generality of our results.
Amershi and others~\cite{amershi} focus on software engineering teams with ML components at Microsoft.
Experts consistently mention challenges around data availability and reusability. The authors synthesize their findings into three major differences between engineering traditional and ML-enabled software: \textit{Data discovery and management}, \textit{Customization and Reuse}, and  \textit{ML Modularity}. 
While the challenges around data clearly correspond to several of our findings (\textcolor{patterncolor}{AP4-5}), the picture for customization and reuse is more complex. The authors focused solely on ML models' reusability for which we could not find any anti-pattern. However, we found additional issues of reusability of features and infrastructure, leading to redundant development (\textcolor{patterncolor}{AP7-9}). Finally, the authors describe challenges for close entangling between model teams. We also find anti-patterns related to organizational silos between model and product team as well as between data producer and consumer (\textcolor{patterncolor}{AP1-3}, \textcolor{patterncolor}{AP4-6}).

Kim and others~\cite{kim2018} investigate the role, work, and background of data scientists also at Microsoft. Their findings motivate further research in different directions, one being centralized data and standardized nomenclature, a recommendation that was also made by our speakers (\textcolor{recommendcolor}{R6-7}). Arpteg and others~\cite{arpteg2018} investigate challenges of deep learning within seven projects and find organizational challenges to be one of three main categories. Challenges include tensions between product management and data scientists and different values between engineers and data scientists, an observation that was frequently shared by the MLOps community (\textcolor{patterncolor}{AP11}, \textcolor{causecolor}{C1}, \textcolor{causecolor}{C6}). 

Consistent with Nahar and others~\cite{nahar}, we see the clash of cultures (\textcolor{causecolor}{C1}) and a lack of engineering competence in data scientists (\textcolor{causecolor}{C2}) in combination with organizational siloing as root causes for tensions between teams. We share several observations for anti-patterns (\textcolor{patterncolor}{AP2}, \textcolor{patterncolor}{AP4-6}) as they also studied socio-technical aspects. In addition, we contextualized some of their findings not as anti-patterns, but as causes to challenges (e.g. \textcolor{causecolor}{C14}, \textcolor{causecolor}{C16}), which provides a new perspective on which practitioners can act on.

For instance, the authors report back that hiring people with mixed skills in engineering and data science seems beneficial, but often absent. We were able to complement this observation by describing anti-patterns and tracing their causes back to unclear roles and titles (\textcolor{causecolor}{C8}) and a general skill shortage on the market (\textcolor{causecolor}{C10}), which amplifies the struggles of uneducated hirers (\textcolor{causecolor}{C9}) to recruit the right staff.

Compared to other papers, our sample holds unique characteristics (i.e., mainly leaders and managers with business and technical expertise). This way, we can abstract from single teams to the whole organization and make novel findings related to lack of communication and leadership. Specifically, these findings relate to missing decision processes within teams (\textcolor{causecolor}{C12}) that finally lead to unmaintainable projects (\textcolor{patterncolor}{AP14}), and a lack of product management (\textcolor{causecolor}{C14-16}) that enforces endless iterations on proof-of-concepts with meaningless metrics (\textcolor{patterncolor}{AP15}).

As a notable methodological contribution, we have shown how practitioner communities can be leveraged to provide a valuable source of qualitative data next to time-consuming (semi-)structured interviews, especially in areas where practitioners are hard to recruit.

The presented anti-patterns can be starting points for future research endeavours. In this regard, a notable open question is whether the found anti-patterns are specific to the development of ML-enabled systems or are, in fact, artifacts of more general socio-technical challenges in software engineering, which at least partly, have been studied before~\cite{tamburri2015social, cataldo2008socio}.

Finally, this paper is not exclusively targeting academics, but enables transfer to industry by providing clear actionables for practitioners. We have gathered a list of 17 anti-patterns that managers and organizational leaders can use to spot socio-technical issues. The identified causes and suggested recommendations help practitioners to act on these anti-patterns. Overall, our findings call for more research on development processes, guidelines, and general organization structures to support the development of ML-enabled software.

\section*{Acknowledgment}

The authors work has been supported by the Federal Ministry of Education and Research of Germany and by the S\"achsische Staatsministerium f\"ur Wissenschaft Kultur und Tourismus in the program Center of Excellence for AI-research "Center for Scalable Data Analytics and Artificial Intelligence Dresden/Leipzig", project identification number: ScaDS.AI. 
Siegmund's work has been funded by the German Research
Foundation (SI 2171/2-2).


\bibliographystyle{IEEEtran}
\bibliography{ref}
\centering{\subsection*{\scshape{Meetups}}}

\begin{footnotesize}
\begin{itemize}
\item[{[M3]}] MLOps.community ``Hierarchy of MLOps Needs'', \textit{YouTube}, Apr 2, 2020. \url{https://www.youtube.com/watch?v=MRES5IxVnME}. 
\item[{[M5]}] MLOps.community ``High Stakes ML: Active Failures, Latent Factors'', \textit{YouTube}, Apr 16, 2020. \url{https://www.youtube.com/watch?v=9g4deV1uNZo}. 
\item[{[M10]}] MLOps.community ``MLOps - The Blind Men and the Elephant'', \textit{YouTube}, May 11, 2020. \url{https://www.youtube.com/watch?v=RTBq7e3FhEw}. 
\item[{[M11]}] MLOps.community ``Machine Learning at Scale in Mercadolibre'', \textit{YouTube}, May 15, 2020. \url{https://www.youtube.com/watch?v=ypySVdT9U7Q}. 
\item[{[M17]}] MLOps.community ``The Challenges of ML Operations and How Hermione Helped on the Way'', \textit{YouTube}, Jun 11, 2020. \url{https://www.youtube.com/watch?v=cDRXLqKJ6I0}. 
\item[{[M21]}] MLOps.community ``Deep Dive on Paperspace Tooling'', \textit{YouTube}, Jul 6, 2020. \url{https://www.youtube.com/watch?v=qMHUE5b1Ee4}. 
\item[{[M22]}] MLOps.community ``Feature Stores: An Essential Part of the ML Stack to Build Great Data'', \textit{YouTube}, Jul 13, 2020. \url{https://www.youtube.com/watch?v=IjO8VUCIZxc}. 
\item[{[M29]}] MLOps.community ``Scaling Machine Learning Capabilities in Large Organizations'', \textit{YouTube}, Aug 11, 2020. \url{https://www.youtube.com/watch?v=aeYnfU26WGk}. 
\item[{[M30]}] MLOps.community ``Path to Production and Monetizing Machine Learning'', \textit{YouTube}, Aug 18, 2020. \url{https://www.youtube.com/watch?v=voO0B0_BsuQ}. 
\item[{[M35]}] MLOps.community ``Bring Your On-Prem ML Use Cases to Production on the Google Cloud using Kubeflow'', \textit{YouTube}, Sep 28, 2020. \url{https://www.youtube.com/watch?v=JZAc_yZgByg}. 
\item[{[M36]}] MLOps.community ``Moving Deep Learning from Research to Production with Determined and Kubeflow'', \textit{YouTube}, Oct 2, 2020. \url{https://www.youtube.com/watch?v=-UymDRk5ISY}. 
\item[{[M37]}] MLOps.community ``When You Say Data Scientist Do You Mean Data Engineer? Lessons Learned From Startup Life'', \textit{YouTube}, Oct 12, 2020. \url{https://www.youtube.com/watch?v=v2HzCcAT1t8}. 
\item[{[M43]}] MLOps.community ``The Current MLOps Landscape'', \textit{YouTube}, Nov 23, 2020. \url{https://www.youtube.com/watch?v=i6HZ2vjFLIs}. 
\item[{[M44]}] MLOps.community ``Human-centric ML Infrastructure: A Netflix Original'', \textit{YouTube}, Dec 14, 2020. \url{https://www.youtube.com/watch?v=TzRNZO2E-eM}. 
\item[{[M45]}] MLOps.community ``How To Move From Barely Doing BI to Doing AI - Building A Solid Data Foundation'', \textit{YouTube}, Dec 20, 2020. \url{https://www.youtube.com/watch?v=0sAyemr6lzQ}. 
\item[{[M46]}] MLOps.community ``Real-time Feature Pipelines, A Personal History'', \textit{YouTube}, Jan 8, 2021. \url{https://www.youtube.com/watch?v=M1p1uJNbHUg}. 
\item[{[M47]}] MLOps.community ``ProductizeML: Assisting Your Team to Better Build ML Products'', \textit{YouTube}, Jan 18, 2021. \url{https://www.youtube.com/watch?v=B7_xPTQtZIE}. 
\item[{[M49]}] MLOps.community ``Machine Learning Design Patterns for MLOps'', \textit{YouTube}, Feb 2, 2021. \url{https://www.youtube.com/watch?v=vH7UFZZdja8}. 
\item[{[M54]}] MLOps.community ``Product Management in Machine Learning'', \textit{YouTube}, Mar 5, 2021. \url{https://www.youtube.com/watch?v=Sl7WrlbXf9E}. 
\item[{[M55]}] MLOps.community ``How to Avoid Suffering in Mlops/Data Engineering Role'', \textit{YouTube}, Mar 12, 2021. \url{https://www.youtube.com/watch?v=7L1W6Y1G-sI}. 
\item[{[M56]}] MLOps.community ``Operationalizing Machine Learning at a Large Financial Institution'', \textit{YouTube}, Mar 19, 2021. \url{https://www.youtube.com/watch?v=vrvagiFVzI4}. 
\item[{[M57]}] MLOps.community ``A Missing Link in the ML Infrastructure Stack'', \textit{YouTube}, Mar 26, 2021. \url{https://www.youtube.com/watch?v=16EnaNnHzOU}. 
\item[{[M58]}] MLOps.community ``Model Watching: Keeping Your Project in Production'', \textit{YouTube}, Apr 4, 2021. \url{https://www.youtube.com/watch?v=7HJ5x-DglLE}. 
\item[{[M59]}] MLOps.community ``MLOps Community 1 Year Anniversary!'', \textit{YouTube}, Apr 9, 2021. \url{https://www.youtube.com/watch?v=Wmyvj1WYxUE}. 
\item[{[M60]}] MLOps.community ``Deploying Machine Learning Models at Scale in Cloud'', \textit{YouTube}, Apr 16, 2021. \url{https://www.youtube.com/watch?v=i3U0gkHX24s}. 
\item[{[M61]}] MLOps.community ``From Idea to Production ML'', \textit{YouTube}, May 3, 2021. \url{https://www.youtube.com/watch?v=WvwclqkEEpE}. 
\item[{[M62]}] MLOps.community ``Law of Diminishing Returns for Running AI Proof-of-Concepts'', \textit{YouTube}, May 10, 2021. \url{https://www.youtube.com/watch?v=j09xbtudJgs}. 
\item[{[M64]}] MLOps.community ``Your Model is Not An Island: Operationalize Machine Learning at Scale with MLOps'', \textit{YouTube}, May 25, 2021. \url{https://www.youtube.com/watch?v=o4NN_Ll4I8o}. 
\item[{[M65]}] MLOps.community ``Common Mistakes in the AI Development Process'', \textit{YouTube}, Jun 1, 2021. \url{https://www.youtube.com/watch?v=OSqb4pmzaWI}. 
\item[{[M68]}] MLOps.community ``Project/Product Management for MLOps'', \textit{YouTube}, Jun 21, 2021. \url{https://www.youtube.com/watch?v=G3gOz_7RBfw}. 
\item[{[M69]}] MLOps.community ``Engineering MLOps'', \textit{YouTube}, Jun 28, 2021. \url{https://www.youtube.com/watch?v=UhoEJxG0duc}. 
\item[{[M75]}] MLOps.community ``I Don't Really Know What MLOps is, but I Think I'm Starting to Like it'', \textit{YouTube}, Aug 20, 2021. \url{https://www.youtube.com/watch?v=VMK5jVT9Rk0}. 
\item[{[M84]}] MLOps.community ``MLOps at Volvo Cars'', \textit{YouTube}, Nov 8, 2021. \url{https://www.youtube.com/watch?v=VzgomadGo1g}. 
\item[{[M92]}] MLOps.community ``Just Build It! Tips for Making ML Engineering and MLOps Real'', \textit{YouTube}, Jan 14, 2022. \url{https://www.youtube.com/watch?v=l1uhE9fEfo8}. 
\item[{[M94]}] MLOps.community ``Trustworthy Data for Machine Learning'', \textit{YouTube}, Feb 21, 2022. \url{https://www.youtube.com/watch?v=k98bPYlZXds}. 
\item[{[M95]}] MLOps.community ``Applications of Data Science'', \textit{YouTube}, Mar 14, 2022. \url{https://www.youtube.com/watch?v=rc8zLY15WZU}. 

\end{itemize}
\end{footnotesize}
\centering{\subsection*{\scshape{Coffee Sessions}}}
\begin{footnotesize}
\begin{itemize}
    \item[{[CS6]}] MLOps.community ``Continuous Integration for ML'', \textit{YouTube}, Aug 10, 2020. \url{https://www.youtube.com/watch?v=L98VxJDHXMM}. 
\item[{[CS13]}] MLOps.community ``How to Choose the Right Machine Learning Tool: A Conversation'', \textit{YouTube}, Oct 15, 2020. \url{https://www.youtube.com/watch?v=mmTCGkm3ZoQ}. 
\item[{[CS18]}] MLOps.community ``Luigi in Production'', \textit{YouTube}, Premiered Nov 9, 2020. \url{https://www.youtube.com/watch?v=ShBod1yXUeg}. 
\item[{[CS19]}] MLOps.community ``Data Observability: The Next Frontier of Data Engineering'', \textit{YouTube}, Nov 23, 2020. \url{https://www.youtube.com/watch?v=IMyI5eKQxMI}. 
\item[{[CS20]}] MLOps.community ``Monzo Machine Learning Case Study'', \textit{YouTube}, Dec 7, 2020. \url{https://www.youtube.com/watch?v=EyLGKmPAZLY}. 
\item[{[CS21]}] MLOps.community ``A Conversation with Seattle Data Guy'', \textit{YouTube}, Dec 8, 2020. \url{https://www.youtube.com/watch?v=o-YXAq9vii4}. 
\item[{[CS23]}] MLOps.community ``SRE for ML Infra'', \textit{YouTube}, Dec 22, 2020. \url{https://www.youtube.com/watch?v=Fu87cHHfOE4}. 
\item[{[CS25]}] MLOps.community ``Most Underrated MLOps Topics'', \textit{YouTube}, Jan 12, 2021. \url{https://www.youtube.com/watch?v=ZNxNrqwRxlc}. 
\item[{[CS26]}] MLOps.community ``Feature Store Master Class'', \textit{YouTube}, Jan 19, 2021. \url{https://www.youtube.com/watch?v=-TGp2qKz8tA}. 
\item[{[CS27]}] MLOps.community ``Practical MLOps'', \textit{YouTube}, Jan 26, 2021. \url{https://www.youtube.com/watch?v=GvAyV8m8ICI}. 
\item[{[CS29]}] MLOps.community ``Culture and Architecture in MLOps'', \textit{YouTube}, Feb 8, 2021. \url{https://www.youtube.com/watch?v=uV676_YLP98}. 
\item[{[CS34]}] MLOps.community ``Machine Learning at Atlassian'', \textit{YouTube}, Apr 12, 2021. \url{https://www.youtube.com/watch?v=MI0hqyYSO3c}. 
\item[{[CS35]}] MLOps.community ``War Stories Productionising ML'', \textit{YouTube}, Apr 19, 2021. \url{https://www.youtube.com/watch?v=DmL_FncITII}. 
\item[{[CS38]}] MLOps.community ``Organisational Challenges of MLOps'', \textit{YouTube}, May 7, 2021. \url{https://www.youtube.com/watch?v=xe3-ImbPkT0}. 
\item[{[CS39]}] MLOps.community ``MLOps: A Leader's Perspective'', \textit{YouTube}, May 18, 2021. \url{https://www.youtube.com/watch?v=LoKMLW1v4EY}. 
\item[{[CS40]}] MLOps.community ``Scaling AI in Production'', \textit{YouTube}, May 21, 2021. \url{https://www.youtube.com/watch?v=iM1tRulj8Xc}. 
\item[{[CS43]}] MLOps.community ``Maturing Machine Learning in Enterprise'', \textit{YouTube}, Jun 15, 2021. \url{https://www.youtube.com/watch?v=kfm3Iozxj8I}. 
\item[{[CS44]}] MLOps.community ``Autonomy vs. Alignment: Scaling AI Teams to Deliver Value'', \textit{YouTube}, Jun 30, 2021. \url{https://www.youtube.com/watch?v=Gr69acrT8HE}. 
\item[{[CS46]}] MLOps.community ``What We Learned from 150 Successful ML-enabled Products at Booking.com.'', \textit{YouTube}, Jul 13, 2021. \url{https://www.youtube.com/watch?v=7MEcm3zINDw}. 
\item[{[CS49]}] MLOps.community ``Aggressively Helpful Platform Teams'', \textit{YouTube}, Aug 10, 2021. \url{https://www.youtube.com/watch?v=az8lXG9v4uo}. 
\item[{[CS56]}] MLOps.community ``A Few Learnings from Building a Bootstrapped MLOps Services Startup'', \textit{YouTube}, Sep 27, 2021. \url{https://www.youtube.com/watch?v=VqMGn6CyRpA}. 
\item[{[CS62]}] MLOps.community ``MLOps from Scratch'', \textit{YouTube}, Premiered Nov 9, 2021. \url{https://www.youtube.com/watch?v=3Fa6uzHxTkQ}. 
\item[{[CS65]}] MLOps.community ``The Future of Data Science Platforms is Accessibility'', \textit{YouTube}, Premiered Nov 30, 2021. \url{https://www.youtube.com/watch?v=7Lmk1f6KV1I}. 
\item[{[CS67]}] MLOps.community ``ML Stepping Stones: Challenges \& Opportunities for Companies'', \textit{YouTube}, Dec 9, 2021. \url{https://www.youtube.com/watch?v=Mk30BuUtH2Y}. 
\item[{[CS69]}] MLOps.community ``Building for Small Data Science Teams'', \textit{YouTube}, Premiered Dec 20, 2021. \url{https://www.youtube.com/watch?v=yAsPfhI5Jd8}. 
\item[{[CS73]}] MLOps.community ``On Structuring an ML Platform 1 Pizza Team'', \textit{YouTube}, Premiered Jan 10, 2022. \url{https://www.youtube.com/watch?v=66A72NgSfeE}. 
\item[{[CS75]}] MLOps.community ``Towards Observability for ML Pipelines'', \textit{YouTube}, Jan 21, 2022. \url{https://www.youtube.com/watch?v=V8U1HksRr_k}. 
\item[{[CS81]}] MLOps.community ``Machine Learning from the Viewpoint of Investors'', \textit{YouTube}, Feb 14, 2022. \url{https://www.youtube.com/watch?v=pjZss-fnOac}. 
\item[{[CS85]}] MLOps.community ``Continuous Deployment of Critical ML Applications'', \textit{YouTube}, Mar 8, 2022. \url{https://www.youtube.com/watch?v=L8WQBYCRaGc}. 
\item[{[CS86]}] MLOps.community ``Building ML/Data Platform on Top of Kubernetes'', \textit{YouTube}, Premiered Mar 12, 2022. \url{https://www.youtube.com/watch?v=u1ggSj0OwMU}. 
\item[{[CS87]}] MLOps.community ``Don't Listen Unless You Are Going to Do ML in Production'', \textit{YouTube}, Mar 17, 2022. \url{https://www.youtube.com/watch?v=eDFCZNZnN-Q}. 
\item[{[CS88]}] MLOps.community ``ML Platform Tradeoffs and Wondering Why to Use Them'', \textit{YouTube}, Mar 28, 2022. \url{https://www.youtube.com/watch?v=DQM-Jue-QiE}. 
\item[{[CS89]}] MLOps.community ``A Journey in Scaling AI'', \textit{YouTube}, Mar 31, 2022. \url{https://www.youtube.com/watch?v=Vj9Qr9H8C8s}. 
\item[{[CS90]}] MLOps.community ``Bringing Audio ML Models into Production'', \textit{YouTube}, Premiered Apr 3, 2022. \url{https://www.youtube.com/watch?v=fGKZljg_SQI}. 
\item[{[CS93]}] MLOps.community ``Model Monitoring in Practice: Top Trends'', \textit{YouTube}, Apr 14, 2022. \url{https://www.youtube.com/watch?v=xnNtq3Swfk4}. 
\item[{[CS94]}] MLOps.community ``Traversing the Data Maturity Spectrum: A Startup Perspective'', \textit{YouTube}, Apr 20, 2022. \url{https://www.youtube.com/watch?v=vZ96dGM3l2k}. 
\item[{[CS95]}] MLOps.community ``MLOps as Tool to Shape Team and Culture'', \textit{YouTube}, Apr 25, 2022. \url{https://www.youtube.com/watch?v=OAB-fu9ylZo}. 

\end{itemize}
\end{footnotesize}

\end{document}